\documentclass[aps,superscriptaddress,showpacs,floatfix,nobibnotes]{revtex4}

\usepackage{subfigure}
\usepackage{epsfig}
\usepackage{epstopdf}
\usepackage{color}
\usepackage{graphicx}
\usepackage{longtable}
\usepackage{CJK}
\usepackage{amsmath,bbold}
\usepackage{float}

\ifx\pdfoutput\undefined
\usepackage[dvips]{graphicx}
\usepackage[dvipdfm,
  pdfstartview=FitH,
     CJKbookmarks=true,
          bookmarksnumbered=true,
          bookmarksopen=true,
         colorlinks=true,
          pdfborder=002,
          citecolor=blue%
           ]{hyperref}
\AtBeginDvi{} 
\else
\usepackage[
          pdfstartview=FitH,
          CJKbookmarks=true,
          bookmarksnumbered=true,
           bookmarksopen=ture,
           colorlinks=true,
           citecolor=blue
           ]{hyperref}

\begin{document}

\begin{CJK*}{GB}{}

\title{Exact solution of the two-axis countertwisting Hamiltonian for the half-integer $J$ case\\
\vskip .3cm
{\small Feng Pan$^{1,3}$, Yao-Zhong Zhang$^{2,4}$, and Jerry P. Draayer$^{3}$} }

\address{Department of Physics, Liaoning Normal University, Dalian 116029, P. R. China\\
$^{2}$School of Mathematics and Physics, The University of Queensland, Brisbane, Qld 4072, Australia\\
$^{3}$Department of Physics and Astronomy, Louisiana State University, Baton Rouge, Louisiana 70803-4001, USA\\
$^{4}$CAS Key Laboratory of Theoretical Physics, Institute of Theoretical Physics\\
      Chinese Academy of Sciences, Beijing 100190, P. R. China
}

\date{\today}

\begin{abstract}
Bethe ansatz solutions of the two-axis countertwisting Hamiltonian for any (integer and half-integer) $J$ are derived
based on the Jordan-Schwinger (differential) boson realization of the $SU(2)$ algebra after desired Euler rotations,
where $J$ is the total angular momentum quantum number of the system.
It is shown that solutions to the Bethe ansatz equations can be obtained as zeros of the extended Heine-Stieltjes polynomials.
Two sets of solutions, with solution number being $J+1$ and $J$ respectively when $J$ is an integer and $J+1/2$ each
when $J$ is a half-integer, are obtained. Properties of the zeros of the related extended Heine-Stieltjes polynomials
for half-integer $J$ cases are discussed. It is clearly  shown that double degenerate level energies for half-integer $J$
are symmetric with respect to the $E=0$ axis. It is also shown that the excitation energies of the `yrast' and
other `yrare' bands can all be asymptotically given by quadratic functions of $J$, especially when $J$ is large.

\end{abstract}


\maketitle

\end{CJK*}

\section{INTRODUCTION}

Spin-squeezed states of both Bose and Fermi many-body systems~\cite{1,11,12,13,14,15,17,18,31,32,33,34}
have been attracting great attention.
Two different models for dynamical generation of spin-squeezed states were proposed
in \cite{1}: the one-axis twisting and the two-axis countertwisting Hamiltonians. As shown in \cite{1},
the latter model gives rise to maximal squeezing with a squeezing angle independent
of system size or evolution time.  Although its experimental implementation has not yet been achieved, it is interesting to
obtain exact solutions of the two-axis countertwisting Hamiltonian for arbitrary total angular momentum $J$.

\vskip .3cm
The Bethe ansatz method provides a powerful tool for generating analytic solutions of a solvable model.
This method was proposed to solve the Hamiltonian of many-spin chain with nearest neighbor interactions~\cite{be},
which was further developed by many researchers (see e.g. \cite{Tak,kor} and references therein).
Many-spin systems with a special type of long-range interactions were also solved exactly by Gaudin in \cite{gau}.
The Gaudin solutions turned out to be equivalent to Richardson's solutions for a nuclear mean-field of the one-body type
plus equal strength pairing interactions among all nucleon pairs~\cite{ric1,ric2}, in which the pairing algebra is the quasi-spin $SU(2)$.
In the Gaudin-Richardson type models~\cite{gau, ric1,ric2}, the one-body terms, which generally can be expressed as
a linear combination of the $SU(2)$ generators, appear in the Hamiltonians. These one-body terms
are essential in the derivation of their algebraic Bethe ansatz solutions.
However, there is no such one-body term in the two-axis countertwisting Hamiltonian.

\vskip .3cm
In fact, the two-axis countertwisting Hamiltonian is equivalent to a special case of the Lipkin-Meshkov-Glick (LMG) model
~\cite{vi1,vi2} after an Euler rotation. As shown previously, similar to other many-spin systems~\cite{be,gau,ric1,ric2},
the LMG model can be solved analytically by using the algebraic Bethe ansatz
\cite{pan, mori}. The same problem can also be solved by using the Dyson boson realization of the
$SU(2)$ algebra~\cite{vi1,vi2,zhang},
of which the solutions may be obtained from the Riccati differential equations~\cite{vi1,vi2}.
Discrete phase analysis of the model with applications to
spin squeezing and entanglement was studied in \cite{mar}. It was also shown \cite{pan1} that
asymmetric rotor Hamiltonian can be solved analytically by using the algebraic Bethe ansatz.
However, as noted in \cite{pan20169}, the procedures used in \cite{pan,pan1} can not be applied
to the two-axis countertwisting Hamiltonian directly mainly due to, as mentioned above, the lack of
one-body terms in the two-axis countertwisting Hamiltonian.

\vskip .3cm
Recently the two-axis countertwisting spin squeezing Hamiltonian was solved exactly for the integer $J$ case \cite{pan20169},
where the algebraic Bethe ansatz is established based on the $SU(1,1)$ algebraic structure.
This is achieved by a combination of the Jordan-Schwinger (differential) boson realization of the $SU(2)$ algebra
and the Fock-Bargmann correspondence. The procedure not only reveals that the Hamiltonian in this case is exactly solvable,
but also shows that the eigenvalue problem can be simplified. Specifically, for integer $J$,
the $2J+1$ dimensional eigenvalue problem is reduced to finding zeros of four independent one-variable polynomials
of ${\rm Int}[J/2]$ or ${\rm Int}[J/2+1]$ order, where ${\rm Int}[p]$ is the integer part of $p$,
due to the underlying hidden $SU(1,1)$ symmetry with four different boson seniority number configurations.

However, as already noted in \cite{pan20169}, the boson mapping procedure there only works for the integer $J$ case.
This can also be seen from the material presented between Eq. (1) and Eq. (7) below, which gives a brief review of the
boson mapping in \cite{pan20169}. In this work, we overcome the shortcomings in the procedure of \cite{pan20169}
by performing a different boson mapping combined with desired Euler rotations and
show that the two-axis countertwisting Hamiltonian can also be exactly solved for half-integers $J$.
One of the keys is introducing two additional parameters in the boson realization so that the Bethe ansatz
can be applied without any constraint on the $J$ values.
Thus our procedure here works for both integer and half-integer $J$ cases of the two-axis countertwisting model.
It will be shown that the $SU(1,1)$ algebraic structure after the suitable transformations is also crucial in the process.
Since solutions for the integer $J$ case have already been shown in \cite{pan20169},
we will mainly focus on solutions for half-integer $J$ cases in this work.
Properties of zeros of the related extended Heine-Stieltjes polynomials for half-integer $J$  will be discussed.
It will be shown that double degenerate level energies for half-integer $J$ cases are symmetric with respect to the $E=0$ axis.
It is also shown that the excitation energies of the `yrast' and other `yrare' bands of the system defined
can all be asymptotically given by quadratic functions of $J$, especially when $J$ is large.

\section{The two-axis countertwisting Hamiltonian}

The two-axis countertwisting Hamiltonian may be written as~\cite{1}
 \begin{equation}\label{1}
 {H}_{\rm TA}={\chi}({J}_{x}J_{y}+J_{y}J_{x}),
 \end{equation}
where $J_{x}$ and $J_{y}$, together with $J_{z}$, are the total angular momentum
operators of the system  and $\chi$ is a constant.
The Hamiltoanian (\ref{1}) is invariant under both parity and
time reversal transformations, namely, it is  $PT$-symmetric.
Due to time-reversal symmetry, level energies of the system are all doubly degenerate when the quantum number $J$ of the
total angular momentum is a half-integer or when $J\rightarrow\infty$ for integer $J$ case~\cite{pan20169}.
As clearly shown in (\ref{1}), the Hamiltonian only contains quadratic terms of $J_{x}$ and $J_{y}$,
for which the Bethe ansatz method~\cite{gau, ric1, ric2} for solving the Gaudin-Richardson model
can not be applied directly.

\vskip .3cm
In our recent paper~\cite{pan20169}, it has been shown that the Hamiltonian
(\ref{1}) can be expressed in terms of the Bargmann variables as
\begin{eqnarray}\label{TA1}
 {H}_{\rm TA}={\chi\over{i}}\left(
 (1+2\delta_{\hat{\nu}_{B}1})
 z_{1}{\partial\over{\partial z_{2}}}-(1+2\delta_{\hat{\nu}_{A}1})z_{2}{\partial\over{\partial z_{1}}}
 +2z_{1}z_{2}({\partial^2\over{\partial z_{2}^2}}-{\partial^2\over{\partial z_{1}^2}})\right),
 \end{eqnarray}
where $\hat{\nu}_{A}$ and $\hat{\nu}_{B}$ are seniority number operator of $A$- and $B$-bosons,
respectively, $i=\sqrt{-1}$, and $z_{j}$ ($j=1,~2$) are Bargmann variables with the mapping:
\begin{equation}\label{TA2}
A^{\dagger\,2}\mapsto z_{1},~~B^{\dagger\,2}\mapsto z_{2}
\end{equation}
after the Jordan-Schwinger realization of the SU(2) algebra with
\begin{equation}\label{T2}
 {J}_{+}=J_{x}+iJ_{y}=A^{\dagger}B,~
 {J}_{-}=J_{x}-iJ_{y}=B^{\dagger}A,~
 {J}_{0}=J_{z}={1\over{2}}(A^{\dagger}A-B^{\dagger}B),
 \end{equation}
where $A^{\dagger}$ ($A$) and $B^{\dagger}$ ($B$) are the boson
creation (annihilation) operators. In (\ref{TA1}), the seniority number ${\nu}_{A}$ and ${\nu}_{B}$
can be taken as $0$ or  $1$, among which the configurations with ${\nu}_{A}={\nu}_{B}$
and those with ${\nu}_{A}\neq{\nu}_{B}$ are related to integer $J$ and half-integer $J$ cases, respectively.
Most notably, the one-body term appears in (\ref{TA1}),
for which the Bethe ansatz method of Gaudin-Richardson becomes applicable.
Nevertheless, not only the Hamiltonian (\ref{TA1}) is non-Hermitian, but also it can not be expressed
in terms of orthonormalized boson modes when ${\nu}_{A}\neq{\nu}_{B}$  as shown in ~\cite{pan20169}.
Therefore, the boson-realization (\ref{TA1}) seems not suitable
for half-integer $J$ case of the Hamiltonian (\ref{1}).
When  ${\nu}_{A}={\nu}_{B}$, which are related to integer $J$ case, however, with further mapping
the Bargmann variables $\{z_{1},~z_{2}\}$ to new boson operators:
$z_{1}\mapsto c^{\dagger}$, $\partial/\partial z_{1}\mapsto c$,
$z_{2}\mapsto d^{\dagger}$, $\partial/\partial z_{2}\mapsto d$,
(\ref{TA1}) may  be written as
\begin{equation}\label{TA3}
 {H}_{\rm TA}={\chi}\left( (1+2\delta_{\hat{\nu} 1})( a^{\dagger}_{1}a_{1}-a^{\dagger}_{2}a_{2})
 +(a^{\dagger 2}_{1}-a^{\dagger 2}_{2})(a^{2}_{1}+a^{2}_{2})\right),
 \end{equation}
where
\begin{equation}\label{a}
 a^{\dagger}_{1}=\sqrt{1\over{2}}(c^{\dagger}+i d^{\dagger}),~~
 a^{\dagger}_{2}=\sqrt{1\over{2}}(c^{\dagger}-i d^{\dagger})
\end{equation}
are two canonical orthonormal boson modes. Though (\ref{TA3}) is still non-Hermitian, its eigenvalues are all real, mainly
because of its equivalence to the original Hamiltonian (\ref{1}) for this case.
Since $\nu_{A}=\nu_{B}\equiv \nu=0$ or $1$, the total angular momentum of the system
should be integer in this case with $J=0,~1,~2,~\cdots$. In other words, the Hamiltonian (\ref{TA3}) is equivalent to
the original one (\ref{1}) only for integer $J$ values. As shown in ~\cite{pan20169},
Hamiltonian (\ref{TA3}), when expressed in terms of $SU(1,1)$ generators, contains a one-body term and can thus be solved
exactly by using the Bethe ansatz method for the four different configurations separately. 

\vskip .3cm
Since the above boson mapping procedure can not be applied to the half-integer $J$ case, 
we need to find an alternative way to solve the problem.
Actually, after rotation around $z$ axis by $-\pi/4$ with
\begin{eqnarray}\label{pi}
J_{x}=\sqrt{1\over{2}} J_{x^{\prime}}-\sqrt{1\over{2}}  J_{y^{\prime}},~~
J_{y}=\sqrt{1\over{2}} J_{y^{\prime}}+\sqrt{1\over{2}}  J_{x^{\prime}},~~
J_{z}=J_{z^{\prime}},
 \end{eqnarray}
(\ref{1}) can be expressed as
\begin{equation}\label{1-1}
 {H}_{\rm TA}^{\prime}={\chi}(J^{2}_{x^{\prime}}-J^{2}_{y^{\prime}}),
 \end{equation}
which is equivalent to a special LMG model whose large $J$ limit was analyzed in \cite{vi1,vi2}
by using the Dyson boson (differential) realization and the corresponding
Riccati differential equations.

\vskip .3cm
Then, after further rotation of the system around $x^{\prime}$ axis by $-\pi/2$ with
\begin{eqnarray}\label{pi2}
J_{x^{\prime\prime}}= J_{x^{\prime}},~~
J_{{y^{\prime\prime}}}=-J_{z^{\prime}},~~
J_{{z^{\prime\prime}}}=J_{y^{\prime}},
 \end{eqnarray}
(\ref{1-1}) becomes
\begin{equation}\label{2b}
H={H}^{\prime\prime}_{\rm TA}={\chi}({J}^{ 2}_{x^{\prime\prime}}-{J}^{2}_{z^{\prime\prime}})=
\chi\left( {1\over{4}}(J_{+}^2+J_{-}^2)+{1\over{2}}C_{2}-{3\over{2}}J^2_{0}\right),
\end{equation}
where $J_{\pm}=J_{x^{\prime\prime}}\pm iJ_{y^{\prime\prime}}$, $J_{0}=J_{z^{\prime\prime}}$, and
$C_{2}={1\over{2}}(J_{+}J_{-}+J_{-}J_{+})+J^{2}_{0}$ is the Casimir operator of the $SU(2)$ generated by $J_{\pm,0}$.
Therefore, up to the Euler rotations, the Hamiltonian (\ref{1}) is equivalent to (\ref{2b}).

\vskip .3cm
Similar to \cite{pan20169}, by using the Jordan-Schwinger realization of $SU(2)$
 \begin{equation}\label{2}
 {J}_{+}=a^{\dagger}b,~~{J}_{-}=b^{\dagger}a, ~~{J}_{0}={1\over{2}}(a^{\dagger}a-b^{\dagger}b),
 \end{equation}
where $a,~b$ and $a^{\dagger},~b^{\dagger}$ are boson annihilation and creation operators,   (\ref{2b}) can be expressed as
\begin{equation}\label{3}
{H}={\chi\over{4}}\left( a^{\dagger 2}b^2 +b^{\dagger 2}a^2+2C_{2}-{3\over{2}}(n_{a}-n_{b})^2
\right).
\end{equation}
Here $n_{a}=a^{\dagger}a$ and $n_{b}=b^{\dagger}b$ are number operators of $a$-bosons and $b$-bosons, respectively.

Introducing two nonzero parameters $c_{1}$ and $c_{2}$, we have the following identities:
\begin{eqnarray}\label{4}\nonumber
&a^{\dagger 2}b^2 +b^{\dagger 2}a^2={1\over{c_{1}c_{2}}}\left(
(c_{1}a^{\dagger 2}+c_{2}b^{\dagger 2})(c_{1}a^2+c_{2}b^2)-c^{2}_{1}n_{a}(n_{a}-1)-c^{2}_{2}n_{b}(n_{b}-1)\right),\\
&n_{a}n_{b}={1\over{c_{1}^2+c^{2}_{2}}}\left(
(c^{2}_{1}n_{a}+c^2_{2}n_{b})(n_{a}+n_{b})-c^{2}_{1}n_{a}^2-c^{2}_{2}n_{b}^2\right).~~~~~~~~~~~~~~~~~~~~~~~~~~~~
\end{eqnarray}
Then (\ref{3}) can be expressed as
\begin{eqnarray}\label{5}\nonumber
{H}&=&{\chi\over{4}}\left\{ {1\over{c_{1}c_{2}}}
(c_{1}a^{\dagger 2}+c_{2}b^{\dagger 2})(c_{1}a^2+c_{2}b^2)+ {1\over{c_{1}c_{2}}}(c^{2}_{1}n_{a}+c^{2}_{2}n_{b})\right.\\ \nonumber
& &-\left( {c_{1}\over{c_{2}}}+{3\over{2}}+{(3+2\lambda)c_{1}^2\over{c_{1}^2+c_{2}^{2}}}-\lambda\right)n^{2}_{a}
-\left( {c_{2}\over{c_{1}}}+{3\over{2}}+{(3+2\lambda)c_{2}^2\over{c_{1}^2+c_{2}^{2}}}-\lambda\right)n^{2}_{b}\\
& &+\left.{3+2\lambda\over{c_{1}^2+c_{2}^{2}}}
(c^{2}_{1}n_{a}+c^{2}_{2}n_{b})(n_{a}+n_{b})-\lambda(n_{a}+n_{b})^2+2C_{2}\right\},
\end{eqnarray}
which is independent of $c_{1}$, $c_{2}$, and $\lambda$, and can be simplified to the desired form
\begin{eqnarray}\label{6}
{H}&=&{\chi\over{4}}\left\{ {1\over{c_{1}c_{2}}}
(c_{1}a^{\dagger 2}+c_{2}b^{\dagger 2})(c_{1}a^2+c_{2}b^2)+ {1\over{c_{1}c_{2}}}(c^{2}_{1}n_{a}+c^{2}_{2}n_{b})\right.\\\nonumber
& &+\left.{3+2\lambda\over{c_{1}^2+c_{2}^{2}}}
(c^{2}_{1}n_{a}+c^{2}_{2}n_{b})(n_{a}+n_{b})-\lambda(n_{a}+n_{b})^2+2C_{2}\right\}
\end{eqnarray}
if
\begin{equation}\label{7}
{c_{1}\over{c_{2}}}+{3\over{2}}+{(3+2\lambda)c_{1}^2\over{c_{1}^2+c_{2}^{2}}}-\lambda=0,~~
{c_{2}\over{c_{1}}}+{3\over{2}}+{(3+2\lambda)c_{2}^2\over{c_{1}^2+c_{2}^{2}}}-\lambda=0
\end{equation}
are satisfied. There are two sets of solutions to (\ref{7}),
$\lambda={3\over{2}},~~c_{1}/c_{2}=-3 \pm 2 \sqrt{2}$,
of which any set may be used for our purpose. In the following, we choose
\begin{equation}\label{constraint}
\lambda={3/{2}},~~~~c_{2}=1, ~~~~c_{1}=-3+ 2 \sqrt{2}.
\end{equation}
Then, (\ref{6}) can be written as
\begin{equation}\label{9}
{H}={\chi\over{4}}\left( {4\over{c_{1}c_{2}}}{S}^{+}S^{-}+({1\over{c_{1}c_{2}}}+
{12J\over{c_{1}^2+c_{2}^{2}}})\left(2S^{0}-{1\over{2}}(c^2_{1}+c^2_{2})\right)-2J(2J-1)\right),
\end{equation}
where use has been made of $n_{a}+n_{b}=2J$ and $C_{2}=J(J+1)$ for a given quantum number of the total angular momentum $J$; moreover
\begin{equation}
S^{\pm}=c_{1}S^{\pm}_{a}+c_{2}S^{\pm}_{b},~~~~~
S^{0}=c_{1}^{2}S^{0}_{a}+c_{2}^{2}S^{0}_{b},
\end{equation}
and
\begin{eqnarray}
&&S^{+}_{a}={1\over{2}}a^{\dagger 2},~~~~
S^{-}_{a}={1\over{2}}a^{2},~~~~S^{0}_{a}={1\over{2}}(a^{\dagger}a+{1\over{2}}),\nonumber\\
&&S^{+}_{b}={1\over{2}}b^{\dagger 2},~~~~S^{-}_{b}={1\over{2}}b^{2},~~~~~S^{0}_{b}={1\over{2}}(b^{\dagger}b+{1\over{2}})
\end{eqnarray}
are respectively $SU_{a}(1,1)$ and $SU_{b}(1,1)$ generators which satisfy the commutation relations
\begin{equation}\label{su11}
[S^{0}_{\rho},~ S^{\pm}_{\rho^{\prime}}]=\delta_{\rho\rho^{\prime}}S^{\pm}_{\rho},~~~
[S^{+}_{\rho},~ S^{-}_{\rho^{\prime}}]=-\delta_{\rho\rho^{\prime}} 2S^{0}_{\rho}.
\end{equation}

It is obvious that the one-body term $S^{0}$ appears in (\ref{9})
after the transformation (\ref{4}) with the constraints (\ref{7}).
Such one-body term is essential in the Bethe ansatz approach described in \cite{gau, ric1,ric2,pan20169,pan,pan1}.
The main difference between the boson realizations (\ref{9}) and (\ref{TA3})
lies in the fact that the former introduces two additional parameters $c_{1}$ and $c_{2}$ with
the constraint (\ref{constraint}), with which the the Bethe ansatz approach described in \cite{gau, ric1,ric2,pan20169,pan,pan1}
can now be applied for both the integer and half-integer $J$ cases.

\vskip .3cm
Similar to what is shown in \cite{pan20169}, (\ref{9}) can now be diagonalized via the Bethe ansatz
\begin{eqnarray}\label{BA}
\vert k,\nu_{a},\nu_{b};\zeta\rangle= S^{+}(w^{(\zeta)}_{1})S^{+}(w^{(\zeta)}_{2})\cdots S^{+}(w^{(\zeta)}_{k})\vert \nu_{a},\nu_{b}\rangle
\end{eqnarray}
with $J=k+{1\over{2}}(\nu_{a}+\nu_{b})$, where
$\vert \nu_{a},\nu_{b}\rangle$ is the lowest weight state of the $SU_{\rho}(1,1)$ for $\rho=a$ and $b$,
satisfying $S^{-}_{\rho}\vert \nu_{a},\nu_{b}\rangle=0$ and $2S^{0}_{\rho}\vert \nu_{a},\nu_{b}\rangle=
\left(\nu_{\rho}+{1\over{2}}\right)\vert \nu_{a},\nu_{b}\rangle$ with $\nu_{\rho}=0$ or $1$,
and
\begin{equation}\label{S}
S^{+}(w)= {c_{1}\over{1-c_{1}^2 w}}S^{+}_{a}+{c_{2}\over{1-c^2_{2} w}}S^{+}_{b}.
\end{equation}
Using the commutation relations (\ref{su11}), it can be proven that
\begin{eqnarray}\label{c1}
[S^{0},~S^{+}(w)]= {c_{1}^3\over{1-c_{1}^2 w}}S^{+}_{a}+{c_{2}^3\over{1-c^2_{2} w}}S^{+}_{b}={1\over{w}}(S^{+}(w)- S^{+}),
\end{eqnarray}
\begin{eqnarray}\label{c2}
[S^{-},~S^{+}(w)]=\Lambda_{0}(w)= {2c^{2}_{1}S^{0}_{a}\over{1-c^{2}_{1}w}}+{2c_{2}^{2}S^{0}_{b}\over{1-c_{2}^2w}},
\end{eqnarray}
\begin{eqnarray}\label{c3}
S^{+}(w_{1},w_{2})=[[S^{-},~S^{+}(w_{1})],~S^{+}(w_{2})]= {2\over{w_{1}-w_{2}}}(
S^{+}(w_{1})-S^{+}(w_{2})).
\end{eqnarray}
With the help of (\ref{c1})-(\ref{c3}), we can directly check that
\begin{eqnarray}\label{10}\nonumber
S^{0}\vert k,\nu_{a},\nu_{b};\zeta\rangle &=&
{1\over{w^{(\zeta)}_{1}}}\left(S^{+}(w^{(\zeta)}_{1})-S^{+})\right)
S^{+}(w^{(\zeta)}_{2})\cdots S^{+}(w^{(\zeta)}_{k})\vert \nu_{a},\nu_{b}\rangle\\
& &+\cdots+S^{+}(w^{(\zeta)}_{1})\cdots S^{+}(w^{(\zeta)}_{k-1})
{1\over{w^{(\zeta)}_{k}}}\left(S^{+}(w^{(\zeta)}_{k})-S^{+})\right)
\vert \nu_{a},\nu_{b}\rangle,
\end{eqnarray}
and
\begin{eqnarray}\label{11}\nonumber
S^{+}S^{-}\vert k,\nu_{a},\nu_{b};\zeta\rangle&=&S^{+}\left(
\overline{\Lambda}_{0}(w^{(\zeta)}_{1})S^{+}(w_{2}^{(\zeta)})\cdots S^{+}(w^{(\zeta)}_{k})\right.\\
& &+\left.\cdots+S^{+}(w^{(\zeta)}_{1})\cdots S^{+}(w^{(\zeta)}_{k-1})\overline{\Lambda}_{0}(w^{(\zeta)}_{k})\right)\vert \nu_{a},\nu_{b}\rangle \nonumber \\
& &+S^{+}\left(S_{+}(w^{(\zeta)}_{1},~w^{(\zeta)}_{2})S^{+}(w^{(\zeta)}_{3})\cdots S^{+}(w^{(\zeta)}_{k})\right.\nonumber \\
& &+S^{+}(w^{(\zeta)}_{1},~w^{(\zeta)}_{3})S^{+}(w^{(\zeta)}_{2})S_{+}(w^{(\zeta)}_{4})\cdots S^{+}(w^{(\zeta)}_{k})\nonumber \\
& &+\cdots+S_{+}(w^{(\zeta)}_{1},~w^{(\zeta)}_{k})S_{+}(w^{(\zeta)}_{2})\cdots S^{+}(w^{(\zeta)}_{k-1})\nonumber \\
& &+\cdots+S_{+}(w^{(\zeta)}_{k},~w^{(\zeta)}_{1})S_{+}(w^{(\zeta)}_{2})\cdots S^{+}(w^{(\zeta)}_{k-1})\nonumber \\
& &+S^{+}(w^{(\zeta)}_{k},~w^{(\zeta)}_{2})S_{+}(w^{(\zeta)}_{3})\cdots S^{+}(w^{(\zeta)}_{k-1})\nonumber \\
& &+\cdots+\left.S^{+}(w^{(\zeta)}_{k},~w^{(\zeta)}_{k-1})S^{+}(w^{(\zeta)}_{2})\cdots S^{+}(w^{(\zeta)}_{k-2})\right)
\vert \nu_{a},\nu_{b}\rangle,
\end{eqnarray}
where $\overline{\Lambda}_{0}(w)={c_{1}^2(\nu_{a}+{1\over{2}})\over{1-c^{2}_{1}w}}+{c^{2}_{2}(\nu_{b}+{1\over{2}})
\over{1-c_{2}^2 w}}$.

Now by means of (\ref{c3})-(\ref{11}), we can prove that the eigen-equation
${H}\vert k,\nu_{a},\nu_{b};\zeta\rangle=E^{(\zeta)}_{k,\nu_{a},\nu_{b}}\vert k,\nu_{a},\nu_{b};\zeta\rangle$
is fulfilled if and only if
\begin{equation}\label{12}
{(\nu_{a}+{1\over{2}})c_{1}^2\over{1-c_{1}^2w^{(\zeta)}_{l}}}
-({1\over{2}}+{6c_{1}c_{2}J\over{c_{1}^2+c_{2}^2}}){1\over{w^{(\zeta)}_{l}}}+{(\nu_{b}+{1\over{2}})
c_{2}^2\over{1-c^{2}_{2}w^{(\zeta)}_{l}}}-
\sum_{j\neq l}{2\over{w^{(\zeta)}_{l}-w^{(\zeta)}_{j}}}
=0~~{\rm for}~~l=1,2,\cdots,~ k.
\end{equation}
The corresponding eigen-energy is given by
\begin{equation}\label{13}
E^{(\zeta)}_{{k,\nu_{a},\nu_{b}}}={\chi\over{4}}\left\{({1\over{c_{1}c_{2}}}+
{12J\over{c_{1}^2+c_{2}^{2}}})
\left(\sum_{l=1}^{k}{2\over{w_{l}^{(\zeta)}}}+c^{2}_{1}\nu_{a}+c_{2}^2\nu_{b}\right)-2J(2J-1)\right\}
\end{equation}
with $J=k+{1\over{2}}(\nu_{a}+\nu_{b})$, where $k$ is the number of boson pairs,
while $\nu_{a}+\nu_{b}$ are the total number of unpaired bosons.
It can be inferred from (\ref{13}) that the spectrum of the Hamiltonian (\ref{9})
is generated from the non-linear boson-pair excitations based on the single-boson excitations,  while
the single-boson excitation energies contribute to the total energy linearly.
Moreover, it is obvious that Eq. (\ref{12}) and the eigenvalues (\ref{13})
are invariant under the simultaneous interchanges  $\nu_{a}\leftrightarrow\nu_{b}$ and $c_{1}\leftrightarrow c_{2}$.
Actually, the eigenvalues of (\ref{5}) for $\{\nu_{a}=1, ~\nu_{b}=0\}$ should be the same as
those for $\{\nu_{a}=0, ~\nu_{b}=1\}$ since the Hamiltonian (\ref{5}) and the constraints (\ref{7})  are all invariant under the permutation of
$c_{1}$ and $c_{2}$, which clearly shows that each level energy is doubly degenerate
when $J$ is a half-integer. Hence, we only need to solve one of the cases for half-integer $J$. The results of the other
case can be obtained by permuting  the $a$-bosons with the $b$-bosons.
In addition, as shown previously \cite{3,4,5}, though the eigenstates provided in (\ref{BA}) are not normalized, they are always orthogonal with
\begin{equation}
 \langle k^{\prime},\nu^{\prime}_{a},\nu^{\prime}_{b};
 \zeta^{\prime}\vert k,\nu_{a},\nu_{b};\zeta\rangle=({\cal N}(k,\zeta;\nu_{a},\nu_{b}))^{-2} \delta_{k k^{\prime}}
 \delta_{\nu_{a} \nu_{a}^{\prime}} \delta_{\nu_{b} \nu_{b}^{\prime}}\delta_{\zeta\zeta^{\prime}},
\end{equation}
where ${\cal N}(k,\zeta;\nu_{a},\nu_{b})$ is the corresponding normalization constant.

\vskip .3cm
According to the Heine-Stieltjes correspondence \cite{3,4,5},
roots of (\ref{12}) are zeros of the extended Heine-Stieltjes polynomials $y_{k}(w)$
of degree $k$  satisfying the following second-order Fuchsian equation:
\begin{equation}\label{15}
y_{k}^{\prime\prime}(w)+\left({\nu_{a}+{1\over{2}}\over{w-c_{1}^{-2}}}+{\nu_{b}+{1\over{2}}\over{w-c_{2}^{-2}}}
+{\gamma_{J}\over{w}}\right)y^{\prime}_{k}(w)+{V(w)\over{w(1-c_{1}^2 w)(1-c_{2}^{2} w) }}y_{k}(w)=0,
\end{equation}
where
\begin{equation}\gamma_{J}=1/2+{6c_{1}c_{2}J\over{c_{1}^2+c_{2}^2}}={1\over{2}}-J,
\end{equation}
and $V(w)$ is a Van Vleck polynomial of degree $1$ determined according to (\ref{15}).
Therefore, the Heine-Stieltjes polynomial approach for solving the Gaudin-Richardson type
equations as shown in \cite{3,4,5} is applicable for solving the Bethe ansatz equations (\ref{12}).
It should be noted that a similar approach for solving the Gaudin-Richardson type equations
has been proposed in Ref. \cite{rr1,rr2}, which however requires the solutions of a set of higher-order differential equations.
Further discussions about the similarities and the differences of the Heine-Stieltjes polynomial approach
and the method shown in \cite{rr1,rr2} may be found in \cite{5,6}.
Since $\nu_{a}+{1\over{2}}$ and $\nu_{b}+{1\over{2}}$ are always real and positive,
while $\gamma_{J}$ is negative except for $J=1/2$, zeros of the extended Heine-Stieltjes polynomials may be complex.
Let the absolute values of the real parts of these complex zeros be arranged as
$\vert {\rm Re}[w_1]\vert< \vert {\rm Re}[w_2]\vert <\cdots< \vert {\rm Re}[w_k]\vert$, which are in the union of
two open intervals:
$\{\vert {\rm Re}[w_{1}]\vert, \vert {\rm Re}[w_{2}]\vert,\cdots, \vert {\rm Re}[w_k]\vert \}\in (0, ~1)\bigcup (1, c_{1}^{-2})$
(noticing that $c_{1}^{-2}>c_{2}^{-2}=1$).

\begin{figure}[H]
\begin{center}
\includegraphics[scale=0.39]{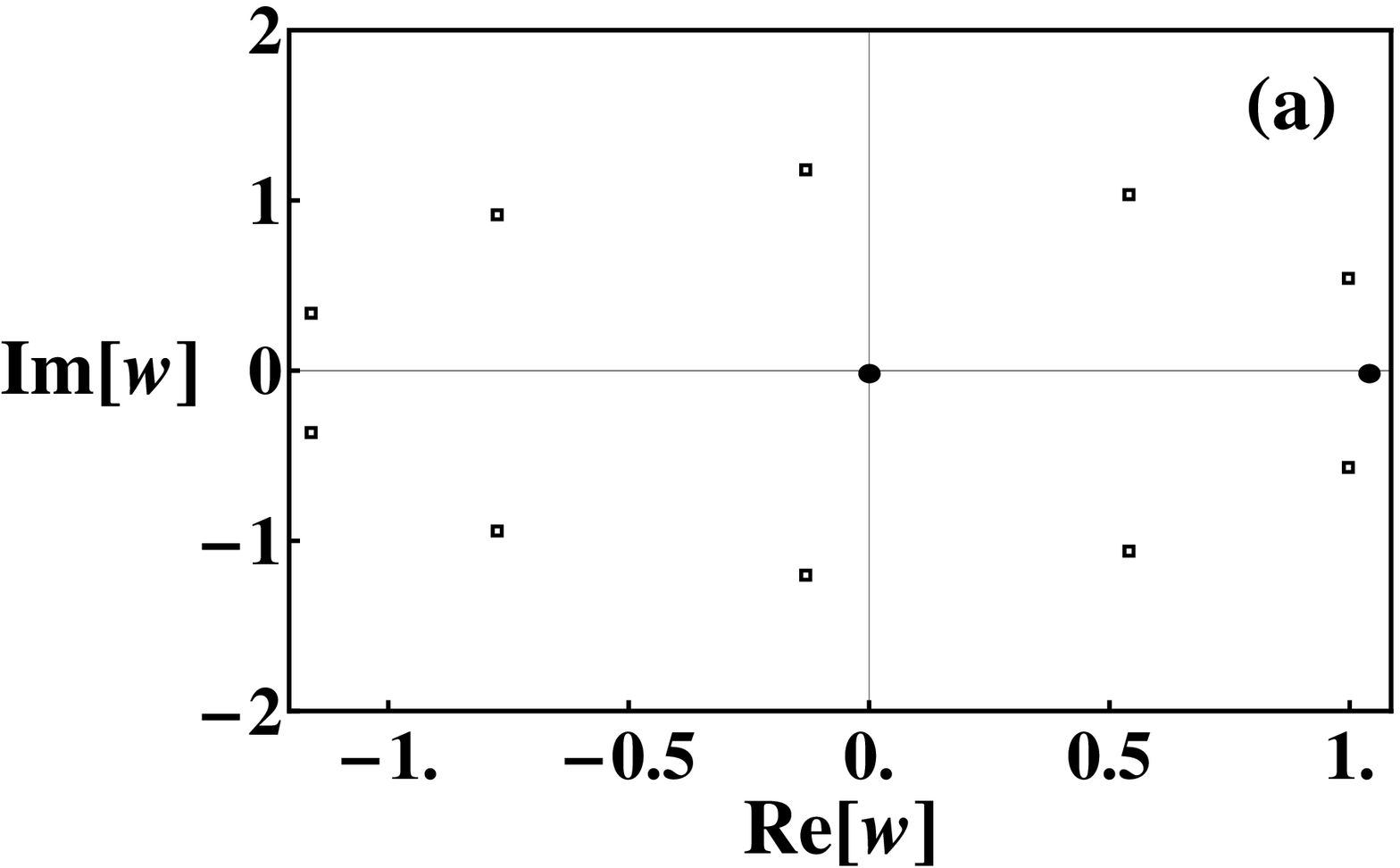}~~
\includegraphics[scale=0.4]{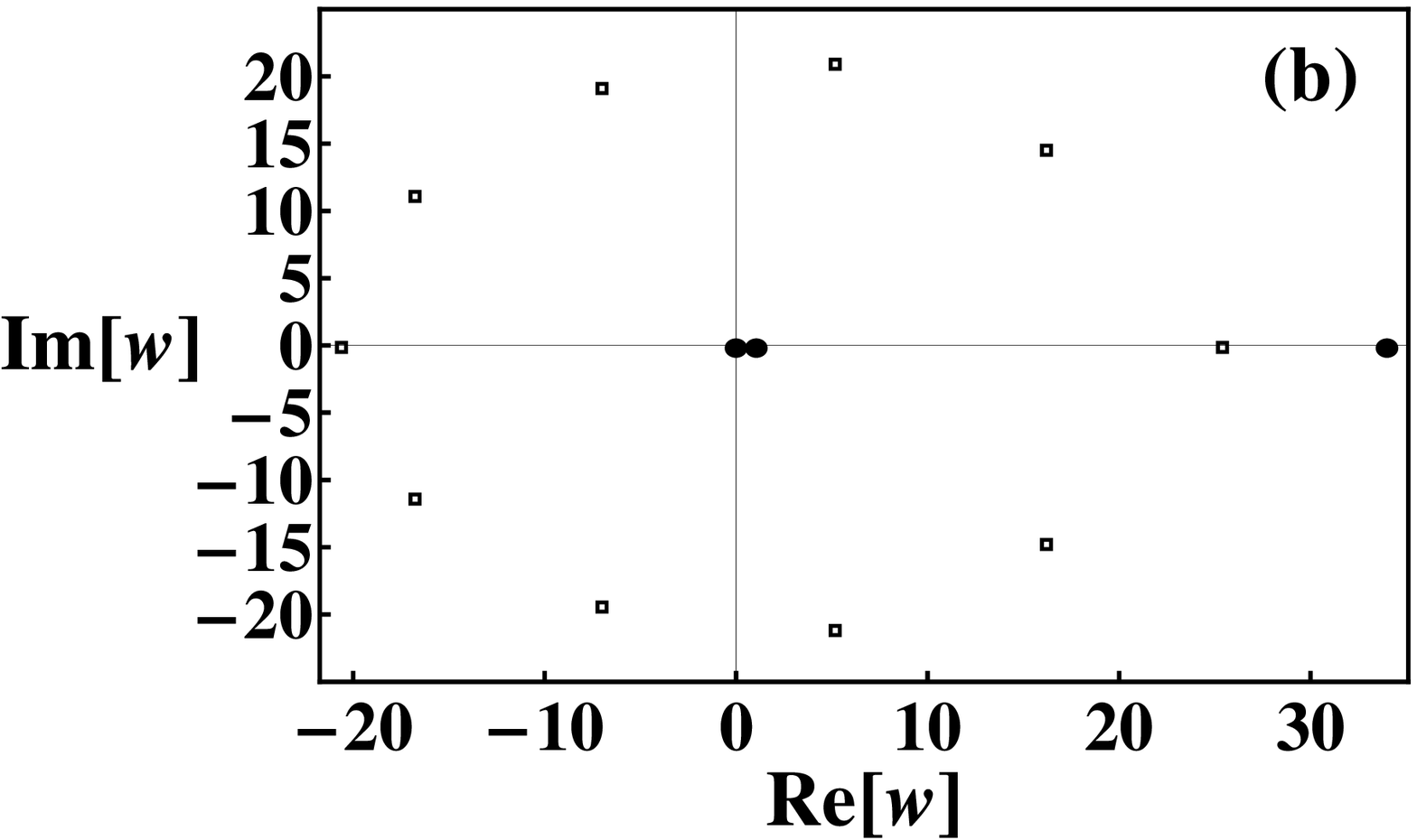}
\caption{The two possible configurations of $10$ zeros
of the extended Heine-Stieltjes polynomial $y_{10}^{(\zeta)}(w)$
for $J=21/2$ ($k=10$) case with  $\{\nu_{a}=0,\nu_{b}=1\}$
corresponding to the ground and the first excited states of the system,
where the solid dot at $0$ on the real axis  represents the negative
charge $\gamma_{J}=1/2-J$ when $J\geq 1/2$, those at $1$ and $c_{1}^{-2}$ on the real axis  represent
the  charges $\nu_{b}+{1\over{2}}={3\over{2}}$ and $\nu_{a}+{1\over{2}}={1\over{2}}$, respectively, the open squares represent the possible positions of the $10$ positive
unit charges or zeros of the polynomial,
(\textbf{a}) the positions of the $10$ zeros of $y_{10}^{(\zeta=1)}(w)$
for the ground state with
eigen-energy $E^{(\zeta=1)}_{ 10,1,0}/\chi=-105.948$, in which
the positive charge ${1\over{2}}$ at $c_{1}^{-2}=33.9706$ is not shown, and (\textbf{b})
the positions of the $10$ zeros of $y_{10}^{(\zeta=2)}(w)$ for the first excited state with eigen-energy $E^{(\zeta=2)}_{ 10,1,0}/\chi=-77.8789$.}
\label{f1}
\end{center}
\end{figure}

\vskip .3cm
An electrostatic interpretation of the location of
zeros of $y_{k}(w)$ may be stated as follows. Put two positive fractional charges ${1\over{2}}\nu_{a}+{1\over{4}}$,
${1\over{2}}\nu_{b}+{1\over{4}}$ at $c_{1}^{-2}$ and $1$,
and one negative fractional charge ${1\over{4}}-{J\over{2}}$ at $0$
 along a real axis of a two-dimensional complex plane, respectively,
and allow k positive unit charges to move freely on the entire two-dimensional complex plane.
There are $k+1$ different configurations for the positions
of these $k$ positive unit charges, of which the absolute values of the
real parts $\{\vert {\rm Re}[w^{(\zeta)}_{1}]\vert,\cdots, \vert {\rm Re}[w^{(\zeta)}_k]\vert\}$ with $\zeta=1,2,\cdots, k+1$,
correspond to global minimums of the total electrostatic energy of the system~\cite{3},
though the imaginary parts of these zeros can not be determined beforehand.
The total number of these configurations
is exactly the number of ways to put the $k$ absolute values of the real parts of the complex zeros into the two open intervals,
which is $k+1$. Thus, there are $k+1$ different polynomials $y_{k}(w)$ for given $\{\nu_{a},\nu_{b}\}$.
Since $0\leq \nu_{a},\nu_{b}\leq 1$, for a given integer $J=k$,
there are $k+1$ solutions for the case with $\{\nu_{a}=\nu_{b}=0\}$,
while there are $k$ solutions for the case with $\{\nu_{a}=\nu_{b}=1\}$.
When $J$ is a half-integer with $J=k+1/2$, there are $k+1$ solutions for the case with $\{\nu_{a}=1,\nu_{b}=0\}$ or
$\{\nu_{a}=0,\nu_{b}=1\}$. Hence, the total number of different solutions equals exactly to $2J+1$
for both integer and half-integer $J$ cases, which proves the completeness of
the solutions provided by (\ref{12}) for the Hamiltonian (\ref{2b}).
Fig. \ref{f1} provides two possible configurations of the $10$ roots for $J=21/2$ ($k=10$) case with  $\{\nu_{a}=0,\nu_{b}=1\}$
corresponding to the ground and the first excited states, which clearly shows that
the absolute values of the real parts of the roots indeed fall into
the union of the two open intervals $(0, ~1)\bigcup (1, c_{1}^{-2})$.
It can be observed that the solutions provided by (\ref{12}) are complex in general
in contrast to the integer $J$ case provided in \cite{pan20169}, of which the solutions are always real.

\vskip .3cm
Once the Bethe ansatz equations (\ref{12}) are solved, the eigenstates (\ref{BA}), up to a normalization constant,
can be expressed in terms of the original $a$- and $b$-boson operators as
\begin{equation}\label{estate}
\vert  k,\nu_{a},\nu_{b};\zeta\rangle=
\sum_{\rho=0}^{k}B_{\rho}^{(k)}
c_{1}^{\rho}
a^{\dagger 2k-2\rho} b^{\dagger 2\rho}\vert \nu_{a},\nu_{b}\rangle,
\end{equation}
where $c_{2}=1$ has been used,
\begin{equation}
B_{\rho}^{(k)}=\sum_{q=0}^{k}(-)^{q}S^{(k,\zeta)}_{q}
\sum_{\mu=0}^{{\rm Min}[\rho,k-q]}
\left(\begin{array}{c}
k-q\\
\mu
\end{array}\right)
\left(\begin{array}{c}
q\\
\rho-\mu
\end{array}\right)c_{1}^{-2\mu},
\end{equation}
and
\begin{equation}\label{Sq}
S^{(k,\zeta)}_{0}=1,~~S^{(k,\zeta)}_{q\geq1}=\sum_{1\leq\mu_{1}\neq\cdots\neq\mu_{q}\leq k}w^{(\zeta)}_{\mu_{1}}
\cdots w^{(\zeta)}_{\mu_{q}}
\end{equation}
are symmetric functions in $\{w^{(\zeta)}_{1},\cdots,w^{(\zeta)}_{k}\}$,
which are related to the expansion coefficients of $y_{k}(w)$ when
it is expanded in terms of powers of $w$ \cite{3,4,5}. Thus, when $J=k$, we have

\begin{equation}\label{exp1}
\vert J=k,\zeta\rangle=\left\{
\begin{tabular}{c}
$\sum_{\rho=0}^{k} {B^{(k)}_{\rho}c^{\rho}_{1}\over{((2k-2\rho)!(2\rho)!)^{-{1\over{2}}}}}
\vert J=k, ~M=k-2\rho\rangle~~~~~~~~~~~~~~~{\rm for}~~\nu_{a}=\nu_{b}=0$,\\\\
$\sum_{\rho=0}^{k-1}{B^{(k-1)}_{\rho}c_{1}^{\rho}\over{((2k-2\rho-1)!(2\rho+1)!)^{-{1\over{2}}}}}
\vert J=k, ~M=k-2\rho-1\rangle
~~~{\rm for}~~\nu_{a}=\nu_{b}=1$.\\\\
\end{tabular}
\right.
\end{equation}
When $J=k+1/2$, we have

\begin{equation}\label{exp2}\small
\vert J=k+1/2,\zeta\rangle=\left\{
\begin{tabular}{c}
$\sum_{\rho=0}^{k} {B^{(k)}_{\rho}c^{\rho}_{1}\over{((2k-2\rho+1)!(2\rho)!)^{-{1\over{2}}}}}
\vert J=k+{1\over{2}}, ~M=k-2\rho+{1\over{2}}\rangle~~~~{\rm for}~~\nu_{a}=1,~\nu_{b}=0$,\\\\
$\sum_{\rho=0}^{k}{B^{(k)}_{\rho}c_{1}^{\rho}\over{((2k-2\rho)!(2\rho+1)!)^{-{1\over{2}}}}}
\vert J=k+{1\over{2}}, ~M=k-2\rho-{1\over{2}}\rangle
~~~~{\rm for}~~\nu_{a}=0,~\nu_{b}=1$.\\\\
\end{tabular}
\right.
\end{equation}
It is clear that there are two set of solutions for the half-integer case, instead of four sets for the integer $J$ case
as shown in \cite{pan20169}.

\vskip .3cm

To solve (\ref{12}) more easily, as shown in \cite{3, 4,5,6}
for the extended Heine-Stieltjes polynomials, we may simply write
\begin{equation}\label{16}
y^{(\zeta)}_{k}(w)=\sum_{j=0}^{k}b^{(\zeta)}_{j}w^{j},
\end{equation}
where $\{b^{(\zeta)}_{j}\}$ ($j=0,~1,~\cdots,~k$) are the $\zeta$-th set of the expansion coefficients to be determined.
Substitution of (\ref{16}) into (\ref{15}) yields the equation which determine the corresponding Van Vleck polynomial
as
\begin{equation}\label{17}
V^{(\zeta)}(w)=c_{1}^2 ~k(J-k-{1\over{2}} -\nu_{a}-\nu_{b})w+g^{(\zeta)}_{0}.
\end{equation}
The expansion coefficients $b^{(\zeta)}_{j}$ and $g^{(\zeta)}_{0}$ satisfy
the following three-term recurrence relation:
\begin{equation}\label{18}
(J-j-1/2)(j+1)b^{(\zeta)}_{j+1}+c_{1}^2(k+1-j)(\nu_{a}+\nu_{b}+k+j-J-1/2)b_{j-1}+
j((c_{1}^2+1)(j-1)+\alpha)b_{j}
=g^{(\zeta)}_{0}b^{(\zeta)}_{j}\end{equation}
where $b^{(\zeta)}_{j}=0$ for $j\leq-1$ or $j\geq k+1$ and
$\alpha=c_1^2 (\nu_a -J+1)+\nu_{b}-J+1$. The recurrence (\ref{18}) is
equivalent to the eigenvalue problem
\begin{equation}\label{19}
{\bf F}{\bf b}^{(\zeta)}=g^{(\zeta)}_{0}{\bf b}^{(\zeta)},\end{equation}
where the transpose of  ${\bf b}^{(\zeta)}$ is
related to the expansion coefficients $\{b^{(\zeta)}_{j}\}$ with
$({\bf b^{(\zeta)}})^{\rm T}=\left( b^{(\zeta)}_{0},b^{(\zeta)}_{1},\cdots,b^{(\zeta)}_{k-1},b^{(\zeta)}_{k}\right)$ and
${\bf F}$ is the $(k+1)\times(k+1)$ tridiagonal matrix with entries determined
by (\ref{18}).

\vskip .3cm

In addition, (\ref{16}) can also be written in terms of the zeros
$\{w^{(\zeta)}_{j}\}$ ($j=1,\cdots,k$) of $y^{(\zeta)}_{k}(w)$ as
\begin{equation}\label{20}
y^{(\zeta)}_{k}(w)=\prod_{j=1}^{k}(w-w^{(\zeta)}_{j})=\sum_{q=0}^{k}(-1)^{q} {S}_{q}^{(k,\zeta)}w^{k-q},
\end{equation}
where  ${S}_{q}^{(k,\zeta)}$ is the symmetric function defined in (\ref{Sq}).
Comparing (\ref{20}) with (\ref{16}), we get
\begin{equation}\label{21}
{S}^{(k,\zeta)}_{q}=(-1)^{q} b^{(\zeta)}_{k-q}\end{equation}
if the overall factor of $\{b^{(\zeta)}_{j}\}$ is chosen to be
$b^{(\zeta)}_{k}=1~\forall~\zeta$, which can be used for the
eigenstates (\ref{exp1}) and (\ref{exp2}) to avoid
unnecessary computation of $S^{(k,\zeta)}_{q}$
from $\{w^{(\zeta)}_{1},\cdots,~w^{(\zeta)}_{k}\}$.

\section{Some numerical examples of the solution}

To demonstrate the method and solutions outlined above, in this section, we provide some examples of the solutions of (1) for half-integer cases.
Due to time reversal symmetry, two sets of the solutions with $\{\nu_{a}=0,\nu_{b}=1\}$
and $\{\nu_{a}=1,\nu_{b}=0\}$ are degenerate for any half-integer $J$.
Actually, eigenstates of the second set of solutions
with $\{\nu_{a}=1,\nu_{b}=0\}$ can be obtained from those with
$\{\nu_{a}=0,\nu_{b}=1\}$ by permuting $a$-bosons with $b$-bosons.
When $J=1/2$, the solutions are trivial with $k=0$. The corresponding
eigen-energies are $E^{(\zeta)}_{{0,1,0}}=E^{(\zeta)}_{{0,0,1}}=0$.
When $J\geq3/2$,  all solutions are non-trivial.
In the following, only some non-trivial $k\neq 0$ solutions for half-inter $J$ cases
will be presented.

\begin{table*}[h]
\caption{The Heine-Stieltjes Polynomials $y^{(\zeta)}_{k}(w)$,
 $g^{(\zeta)}_{0}$ of the corresponding Van Vleck Polynomial $V^{(\zeta)}(w)$,
 and the corresponding eigenenergy $E^{(\zeta)}_{ k,\nu_{a},\nu_{b}}/\chi$  of the Hamiltonian (1) for $J\leq5$, where the order of $\zeta$ is arranged according to
 the value of the eigen-energy of (1) for a given set of $\{ k,\nu_{a},\nu_{b}\}$.}
\begin{tabular}{ccccc}
\hline\hline
$J$&$\{k,\zeta; \nu_{a},\nu_{b}\}$ &{$y^{(\zeta)}_{k}(w)$} &$g^{(\zeta)}_{0}$ &$E^{(\zeta)}_{ k,\nu_{a},\nu_{b}}/\chi$ \\
\hline
\hline
3/2 &$\{1, 1; 1,0\}$ &$18.3378 +w$ &$0.0545323$ &$-1.73205$\\
    &$\{1, 2; 1,0\}$ &$-1.85249+w$ &$-0.539814$ &$~~1.73205$\\
    &$\{1, 1; 0,1\}$ &$1.85249 +w$ &$0.539814$ &$-1.73205$\\
    &$\{1, 2; 0,1\}$ &$-18.3378+w$ &$-0.0545323$ &$~~1.73205$\\

5/2 &$\{2, 1; 1,0\}$ &$448.949 +17.8348 w+w^2$&$17.8348 $&$-5.2915$\\
    &$\{2, 2; 1,0\}$ &$-5.82843+2.41421 w+w^2$&$2.41421 $&$0$\\
    &$\{2, 3; 1,0\}$ &$14.9816 -13.0064 w+w^2$&$-13.0064$&$5.2915$\\
    &$\{2, 1; 0,1\}$ &$2.57044 +1.3495 w+w^2$&$1.3495 $&$-5.2915$\\
    &$\{2, 2; 0,1\}$ &$-197.995-14.0711 w+w^2$&$-14.0711$&$0$\\
    &$\{2, 3; 0,1\}$ &$77.0275 -29.4916 w+w^2$&$-29.4916$&$5.2915$\\
7/2 &$\{3, 1; 1,0\}$ &$12572.6 +445.604 w+17.6893 w^2+w^3$&$445.604  $&$-10.8624$\\
    &$\{3, 2; 1,0\}$ &$-13.2801+5.62989 w+2.084 w^2+w^3$&$5.62989 $&$-2.83003$\\
    &$\{3, 3; 1,0\}$ &$113.704 -85.0097 w-8.91243 w^2+w^3$&$-85.0097$&$2.83003$\\
    &$\{3, 4; 1,0\}$ &$-80.9492+97.7069 w-24.5177 w^2+w^3$&$97.7069$&$10.8624$\\
    &$\{3, 1; 0,1\}$ &$3.11805 +1.62364 w+1.204 w^2+w^3$&$1.62364  $&$-10.8624$\\
    &$\{3, 2; 0,1\}$ &$-2951.93-181.093 w-14.4013 w^2+w^3$&$-181.093 $&$-2.83003$\\
    &$\{3, 3; 0,1\}$ &$344.772 -90.4534 w-25.3977 w^2+w^3$&$-90.4534$&$2.83003$\\
    &$\{3, 4; 0,1\}$ &$-484.279+349.521 w-41.003 w^2+w^3$&$349.521  $&$10.8624$\\

9/2 &$\{4, 1; 1,0\}$ &$372346 +12541.9 w+444.373 w^2+17.6295 w^3+w^4$&$12541.9  $&$-18.4421$\\
    &$\{4, 2; 1,0\}$ &$-22.883+9.993 w+3.49161 w^2+1.65038 w^3+w^4$&$9.993$&$-7.47579$\\
    &$\{4, 3; 1,0\}$ &$1154 -873.992 w-101.912 w^2-9.24264 w^3+w^4$&$-873.992  $&$0$\\
    &$\{4, 4; 1,0\}$ &$-434.327+468.213 w-49.1045 w^2-20.1357 w^3+w^4$&$468.213  $&$7.47579$\\
    &$\{4, 5; 1,0\}$ &$479.231 -742.042 w+314.622 w^2-36.1148 w^3+w^4$&$-742.042$&$18.4421$\\

    &$\{4, 1; 0,1\}$ &$3.57655 +1.8561 w+1.37723 w^2+1.14425 w^3+w^4$&$1.8561 $&$-18.4421$\\
    &$\{4, 2; 0,1\}$ &$-58196.6-2827.34 w-176.083 w^2-14.8349 w^3+w^4$&$-2827.34 $&$-7.47579$\\
    &$\{4, 3; 0,1\}$ &$1154 -313.978 w-101.912 w^2-25.7279 w^3+w^4$&$-313.978$&$0 $\\
    &$\{4, 4; 0,1\}$ &$-3066.16+1817.43 w+130.47 w^2-36.6209 w^3+w^4$&$1817.43  $&$7.47579$\\
    &$\{4, 5; 0,1\}$ &$2778.86 -2954.26 w+757.618 w^2-52.6001 w^3+w^4$&$-2954.26$&$18.4421$\\
    \hline\hline
\end{tabular}\label{t1}
\end{table*}

\begin{center}\begin{table}
\centering\textwidth1.038pt\tabcolsep0.032in\fontsize{8.pt}{8.pt}\selectfont
\caption{The same as Table \ref{t1}, but for $J=21/2$.}
\begin{tabular}{ccccc}
\hline\hline
&$\{k,\zeta; \nu_{a},\nu_{b}\}$ &{$y^{(\zeta)}_{k}(w)$} &$g^{(\zeta)}_{0}$ &$E^{(\zeta)}_{ k,\nu_{a},\nu_{b}}/\chi$ \\
\hline
\hline
  &$\small\{10, 1; 1,0\}$ &$3.66943\times10^{14}+1.13715\times10^{13} w+3.54484\times10^{11} w^2+1.11326\times 10^{10} w^3$&$1.13715\times10^{13}$ &$-105.948$\\
  &&$+3.53008\times10^{8} w^4+1.13402\times10^{7} w^5+371120 w^6+w^{10}~$\\
  &$\{10, 2; 1,0\}$ &$-101.618+45.7896 w+15.0023 w^2+7.07498 w^3+4.31317 w^4+2.98896 w^5 $&$\small 45.7896 $&$-77.8789$\\
  &&$+2.22996 w^6+1.74641 w^7+1.4158 w^8+1.17792 w^9+w^{10}~$\\
  &$\{10, 3; 1,0\}$ &$7.80873\times10^{10}-6.86055\times10^{10} w-4.95308\times10^{9} w^2 -2.71706\times10^{8} w^3$ &$ -6.86055\times10^{10}$&$-52.9351$\\
  &&$-1.36749\times10^{7} w^4-676937 w^5-34672.2 w^6-1969.52 w^7-141.08 w^8+w^{10}$\\
  &$\{10, 4; 1,0\}$ &$-106096+132604 w-18511.1 w^2-6740.2 w^3-2078.69 w^4$&$132604$&$-31.2956$\\
  &&$-795.203 w^5-354.007 w^6-166.08 w^7-74.4573 w^8-25.9728 w^9+w^{10}$~~\\
  &$\{10, 5; 1,0\}$ &$2.55952\times10^8-3.98371\times10^8 w+1.24488\times10^8 w^2+1.85329\times10^7 w^3$&$-3.98371\times10^{8}$&$-13.4271$\\
  &&$+1.82955\times10^6 w^4+166734 w^5+16160.7 w^6$\\
  &&$+1698.38 w^7+113.787 w^8-36.3873 w^9+w^{10}$\\
  &$\{10, 6; 1,0\}$ &$-7.7618\times10^{6}+1.38688\times10^7 w-6.12768\times10^6 w^2-197323 w^3$&$1.38688\times10^{7}$&$0$\\
  &&$+117925w^4+39932.7 w^5+10231.3 w^6$\\
  &&$+2303.31 w^7+334.544 w^8-44.2132 w^9+w^{10}$~\\
  &$\{10, 7; 1,0\}$ &$2.16991\times10^7-4.37708\times10^7 w+2.49872\times10^7 w^2-1.87605\times10^6 w^3$&$-4.37708\times10^{7}$&$13.4271 $\\
  &&$-824936 w^4-126530 w^5-9418.12 w^6$\\
  &&$+1486.04 w^7+623.35 w^8-52.0391 w^9+w^{10}$~~\\
  &$\{10, 8; 1,0\}$ &$-1.85276\times10^7+4.30535\times10^7 w-3.19566\times10^7 w^2+6.76683\times10^6 w^3$&$4.30535\times10^{7}$&$31.2956$\\
  &&$+766832 w^4-98749.6 w^5-38784.5 w^6$\\
  &&$-2953.46 w^7+1113.23 w^8-62.4536 w^9+w^{10}$~\\
  &$\{10, 9; 1,0\}$ &$1.87476\times10^7-5.05252\times10^7 w+4.7973\times10^7 w^2-1.78911\times10^7 w^3$&$-5.05252\times10^{7}$&$52.9351$\\
  &&$+1.26629\times10^6 w^4+478782 w^5-17931.1 w^6$\\
  &&$-15528.5 w^7+1867.83 w^8-75.066 w^9+w^{10}$~\\
  &$\{10, 10;1,0\}$ &$-1.87384\times10^7+5.85199\times10^7 w-6.95261\times10^7 w^2+3.86517\times10^7 w^3$&$5.85199\times10^{7}$&$77.8789$\\
  &&$-9.50417\times10^6 w^4+389737 w^5+239926 w^6$\\
  &&$-43240 w^7+2956.95 w^8-89.6043 w^9+w^{10}$\\
  &$\{10, 11;1,0\}$ &$1.87386\times10^7-6.7545\times10^7 w+9.83678\times10^7 w^2-7.47613\times10^7 w^3 $&$-6.7545\times10^{7} $&$105.948$\\
  &&$+3.21533\times10^7 \times w^4-8.01874\times10^6 w^5+1.16068\times10^6 w^6$\\
  &&$-96440.7 w^7+4463.39 w^8-105.964 w^9+w^{10}$\\
  &$\{10, 1; 0,1\}$ &$5.57736 +2.87943 w+2.13913 w^2+1.77814 w^3+1.55429 w^4$&$2.87943 $&$-105.948$\\
  &&$+1.39809 w^5+1.28114 w^6+1.18935 w^7+1.11482 w^8+1.05274 w^9+w^{10}$\\
  &$\{10, 2; 0,1\}$ &$-2.01399\times10^{13}-6.98348\times10^{11} w-2.47089\times10^{10} w^2-8.97212\times10^8 w^3$&$-6.98348\times10^{11}$&$-77.8789$\\
  &&$-3.37243\times10^7 w^4-1.33065\times10^6 w^5-56524.7 w^6+w^{10} $\\
  &$\{10, 3; 0,1\}$ &$26208.8 -10307.7 w-3204.1 w^2-1316.74 w^3-682.365 w^4$&$-10307.7$&$-52.9351$\\
  &&$-392.176 w^5-233.215 w^6-136.404 w^7-73.1981 w^8-29.8456 w^9+w^{10} $\\
  &$\{10, 4; 0,1\}$ &$-1.92898\times10^{10}+1.47483\times10^{10} w+1.2446\times10^9 w^2+8.17213\times10^7 w^3$&$1.47483\times10^{10}$&$-31.2956$\\
  &&$+5.12777\times10^6 w^4+339072 w^5+26091.6 w^6$\\
  &&$+2490.47 w^7+201.344 w^8-42.4581 w^9+w^{10} $\\
  &$\{10, 5; 0,1\}$ &$7.99592\times10^6-8.56477\times10^6 w+788414. w^2+346414 w^3+97032.7 w^4$&$-8.56477\times10^6$&$-13.4271$\\
  &&$+29469.9 w^5+9519.11 w^6+2838.52 w^7+561.275 w^8-52.8726 w^9+w^{10}$\\
  &$\{10, 6; 0,1\}$ &$-2.63673\times10^8+3.43174\times10^8 w-7.64386\times10^7 w^2$&$3.43174\times10^8 $&$0$\\
  &&$-1.5492\times10^7 w^3-2.02576\times10^6 w^4 -232745 w^5-20232.7 w^6$\\
  &&$+996.608 w^7+911.043 w^8-60.6985 w^9+w^{10}$\\
  &$\{10, 7; 0,1\}$ &$9.43159\times10^7-1.44481\times10^8 w+5.09462\times10^7 w^2$&$-1.44481\times10^8$&$13.4271$\\
  &&$+3.57526\times10^6 w^3-667019 w^4-263794 w^5-50627.8 w^6$\\
  &&$-3389.31 w^7+1328.86 w^8-68.5244 w^9+w^{10} $\\
  &$\{10, 8; 0,1\}$ &$-1.10461\times10^8+2.03078\times10^8 w-1.06558\times10^8 w^2+8.32205\times10^6 w^3$&$2.03078\times10^8$&$31.2956$\\
  &&$+3.21702\times10^6 w^4+241118 w^5-55117.7 w^6$\\
  &&$-14317.7 w^7+1990.42 w^8-78.9389 w^9+w^{10} $\\
  &$\{10, 9; 0,1\}$ &$1.09164\times10^8-2.41225\times10^8 w+1.76691\times10^8 w^2-4.32417\times10^7 w^3$&$-2.41225\times10^8 $&$52.9351$\\
  &&$-1.46987\times10^6 w^4+1.15533\times10^6 w^5+89949.6 w^6$\\
  &&$-37410.9 w^7+2952.95 w^8-91.5513 w^9+w^{10} $\\
  &$\{10, 10;0,1\}$ &$-1.09218\times10^8+2.88085\times10^8 w-2.79855\times10^8 w^2+1.20468\times10^8 w^3$&$2.88085\times10^8$&$77.8789$\\
  &&$-1.96771\times10^7 w^4-940918 w^5+675448 w^6$\\
  &&$-80861.7 w^7+4281.74 w^8-106.09 w^9+w^{10} $\\
  &$\{10, 11;0,1\}$ &$1.09217\times10^8-3.4068\times10^8 w+4.22424\times10^8 w^2-2.68684\times10^8 w^3$&$-3.4068\times10^8 $&$105.948$\\
  &&$+9.51901\times10^7 w^4-1.93589\times10^7 w^5+2.28507\times10^6 w^6$\\
  &&$-156404 w^7+6057.88 w^8-122.45 w^9+w^{10}$\\
\hline\hline
\end{tabular}\label{t2}
\end{table}
\end{center}

The Heine-Stieltjes polynomials $y_{k}^{(\zeta)}(w)$
and the corresponding coefficient $g_{0}^{(\zeta)}$ in the Van Vleck polynomials (\ref{17})
up to $J=9/2$ are shown in Table \ref{t1}, while the $J=21/2$ case is provided in Table \ref{t2}.
For any case, it can be verified that the absolute value of the real part
of any zero of $y_{k}^{(\zeta)}(w)$ indeed lies in one of the
intervals $(0, 1)$ and $(1, c_{1}^{-2})$. When a zero is complex, there must be its complex conjugation
as another zero, which is clearly demonstrated in Fig. \ref{f1}. Thus, eigenvalues (\ref{13}) are always real.
The two-axis countertwisting Hamiltonian can be easily diagonalized by brute force
for small value of total angular momentum $J$, i.e. when the number of particles of the system is small.
We can check that the eigen-energies shown in Tables \ref{t1} and \ref{t2} are exactly the same as those
given in Table I of \cite{1509} obtained from the direct $2J+1$ dimensional matrix diagonalization
for the corresponding half-integer $J$ values. 

\vskip .3cm
Furthermore, by using (\ref{21}), the eigen-energies given in (\ref{13}) can also be expressed as
\begin{equation}\label{24}
E^{(\zeta)}_{{k,\nu_{a},\nu_{b}}}={\chi\over{4}}\left\{({1\over{c_{1}c_{2}}}+
{12J\over{c_{1}^2+c_{2}^{2}}})
\left(-2 b_{1}^{(\zeta)}/b^{(\zeta)}_{0}+c^{2}_{1}\nu_{a}+c_{2}^2\nu_{b}\right)-2J(2J-1)\right\}\end{equation}
with $J=k+(\nu_{a}+\nu_{b})/2$, of which the corresponding numerical values
are also provided in the last column of Tables \ref{t1} and \ref{t2}. It is shown in these Tables
that there is an excited state with $E_{{ k,n_{1},n_{2},\nu}}=0$ when $k$ is zero or even for $J=k+{1\over{2}}$,
while all excited energies are non-zero when $k$ is odd, which applies to any $J$.

\vskip .3cm
\begin{figure}[H]
\begin{center}
\includegraphics[scale=0.39]{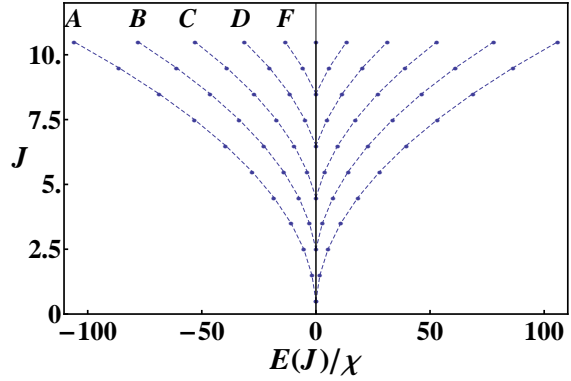}
\caption{The level energies $E(J)/\chi$ denoted by solid dots
up to $J=21/2$, where the dots connected by the dashed lines
belong to the yrast and yrare bands, of which only the first $5$ bands
denoted by $A$ to $F$ and their mirror symmetric ones after the
reflection with $E(J)\rightarrow -E(J)$ are shown. }
\label{f2}
\end{center}
\end{figure}

As demonstrated in Fig. \ref{f2} for level energies of the system
up to $J=21/2$, the level energy distribution for pairs of the double-degenerate levels
is symmetric with respect to $E=0$ axis.
There are ${\rm Int}[k/2]+1$ doubly degenerate levels for odd $k$
or $k/2$ doubly degenerate levels for even $k$ with energies $E_{r}<0$
and the same number of pairs of  doubly degenerate levels with energies  $E_{r}>0$
for $r=1,~2,\cdots, ~{\rm Int}[k/2]+1$ when $k$ is odd ($k/2$ when $k$ is even),
which is consistent with the conclusion  in \cite{pan20169} for integer $J$ cases.

\vskip .3cm
More interestingly, similar to the term used in analyzing energy spectrum of a nucleus~\cite{rev67},
we may define the yrast and other yrare bands of the two-axis countertwisting Hamiltonian.
The yrast band consists of the double-degenerate minimum (ground state)
energies of the system for all possible half-integer $J$,
and the next yrare band consists of the first excitation energies for all possible half-integer $J$, and so on.
For given $k$ with $\nu_{a}=0$ and $\nu_{b}=1$ for example, according to (\ref{24}), these yrast and yrare bands are given by
\begin{equation}\label{25}
E^{(\zeta)}(J)/\chi=-{\left(J^2-({\Omega(\zeta,J)}+{1\over{2}}-{1\over{2c_{1}}})J+
{1\over{2}}({\Omega(\zeta,J)}-{1\over{2c_{1}}})
\right)}\end{equation}
with $\zeta=1,~2,\cdots$, where $\Omega(\zeta,J)={b^{(\zeta)}_{1}/({b^{(\zeta)}_{0}c_{1})}}$
because  $b^{(\zeta)}_{1}$ and $b^{(\zeta)}_{0}$ are also $J$-dependent
with $J=k+1/2$. This provides the level energies of the
yrast band with $\zeta=1$ corresponding to band A in Fig. \ref{f2},
those in the next yrare band with $\zeta=2$ corresponding to band B in Fig. \ref{f2}, and so on.
Since the spectrum is symmetric with respect to the $E=0$ axis,
only level energies with $E^{(\zeta)}(J)<0$ with $\zeta=1,~2,~\cdots$, need to be
analyzed. Though  $\Omega(\zeta,J)$ varies with $J$ and $\zeta$
non-linearly, numerical analysis shows that it is almost a constant
for a given $\zeta$. As clearly shown in Fig. \ref{f2}, $E^{(\zeta)}(J)/\chi$ for a given $\zeta$
uniformly follows a smooth curve described as a quadratic  function of $J$.  For example,
the level energies in the yrast band shown by curve A in Fig. \ref{f2} may be
asymptotically  given by
\begin{equation}\label{26}
E^{(\zeta=1)}(J)/\chi=-J^2+ 0.413J-0.0435\end{equation}
for $J=1/2,~3/2,\cdots$;
the level energies in the next yrare band shown by  curve B in Fig. \ref{f2} may be
asymptotically given by
\begin{equation}\label{27}
E^{(\zeta=2)}(J)/\chi=-J^2+ 3.26J-1.9\end{equation}
for $J=5/2,~7/2,\cdots$, which are more accurate when $J$ is large.

\section{SUMMARY}

In this work, Bethe ansatz solutions of the two-axis countertwisting Hamiltonian for any (integer as well as half-integer) $J$
are derived based on the Jordan-Schwinger (differential) boson realization of the $SU(2)$ algebra after the desired Euler rotations.
It is shown that the solutions to the Bethe ansatz equations can be obtained more easily as zeros of the extended Heine-Stieltjes polynomials.
It is verified that the zeros of the extended Heine-Stieltjes polynomials
may be complex. Though we can not determine the imaginary parts of the zeros
beforehand, the absolute values of the real parts of these complex zeros
are all within the two open intervals $(0,1)$ and $(1,c_{1}^{-2})$.
Since each level is doubly degenerate for half-integer $J$ due to time reversal symmetry, we only need to solve one of the cases.
The results of the other case can be obtained by permuting the
$a$-bosons with the $b$-bosons, which is efficiently helpful
in determining all excited states of the system.
However, unlike the procedure in \cite{pan20169} for integer $J$ case,
in which the $2J+1$ solutions split into four sets of independent solutions,
the procedure presented here gives rise to two sets of solutions
with solution number being $J+1$ and $J$ respectively when $J$ is an integer
and being $J+1/2$ each when $J$ is a half-integer.
It is clearly  shown that double degenerate level energies
for half-integer $J$ case are symmetric with respect to
the $E=0$ axis. It is also shown that the excitation energies
of the `yrast' and other `yrare' bands defined can all be asymptotically given  by
quadratic functions of $J$, especially when $J$ is large.
Our procedure may be used in calculating physical
quantities of the system in order to produce maximal
squeezed spin states of many-particle systems, especially when
the number of particles is large.

\bigskip

\begin{acknowledgments}
{Support from U.S. National Science Foundation
(OCI-0904874, {ACI -1516338}), {U.S. Department
of Energy (DE-SC0005248)}, Southeastern Universities Research Association,
the China-U.S. Theory Institute for Physics with Exotic Nuclei (DE-SC0009971),
National Natural Science Foundation of China (11375080 and 11675071),
Australian Research Council Discovery Project DP140101492,
and LSU--LNNU Joint Research Program (9961) is acknowledged.}

\end{acknowledgments}


\begin{thebibliography}{99}

\bibitem{1} M. Kitagawa and M. Ueda, Phys. Rev. A 47, 5138 (1993).

\bibitem{11} D. J. Wineland, J. J. Bollinger, W. M. Itano, F. L. Moore
and D. J. Heinzen, Phys. Rev. A 46, R6797 (1992).

\bibitem{12} D. J. Wineland, J. J. Bollinger, W. M. Itano and D. J.
Heinzen, Phys. Rev. A 50, R67 (1994).

\bibitem{13} A. S{\o}rensen and K. M{\o}mer, Phys. Rev. Lett. 86, 4431 (2001).

\bibitem{14} J. Hald, J. L. S{\o}rensen, C. Schori and E. S. Polzik, Phys. Rev. Lett. 83, 1319 (1999).

\bibitem{15} I. D. Leroux, M. H. Schleier-Smith and V. Vulet\'{c},
Phys. Rev. Lett. {104}, 073602 (2010).

\bibitem{17} C. D. Hamley, C. S. Gerving, T. M. Hoang, E. M. Bookjans and M. S. Chapman, Nature Phys. 8,  305 (2012).

\bibitem{18} H. Strobel, W. Muessel, D. Linnemann and T. Zibold, Science 345, 424 (2014).

\bibitem{31} V. Meyer, M. A. Rowe, D. Kielpinski, C. A. Sackett,
W. M. Itano, C. Monroe and D. J. Wineland, Phys. Rev. Lett. 86, 5870 (2001).

\bibitem{32} J. Est{\'{e}}ve, C. Gross, A. Weller, S. Giovanazzi and M. K.
Oberthaler, Nature 455, 1216 (2008).

\bibitem{33} J. Appel, P. J. Windpassinger, D. Oblak, U. B. Hoff,
N. Kj{\ae}gaard and E. S. Polzik, PNAS 106, 10960 (2009).

\bibitem{34} M. H. Schleier-Smith, I. D. Leroux  and V. Vulet\'{c},
Phys. Rev. Lett. {104}, 073604 (2010).

\bibitem{be} H. Bethe, Z. Phys. 71, 205 (1931).

\bibitem{Tak} L. Takhtajan, L. Faddeev, J. Sov. Math.
24,  241 (1984).

\bibitem{kor} V. Korepin, N. Bogoliubov, A. Izergin, Quantum Inverse Scattering Method
and Correlation Functions, Cambridge monographs on Mathematical Physics
(Cambridge University Press, 1993).

\bibitem{gau} M. Gaudin, J. Phys. (Paris) 37, 1087 (1976).

\bibitem{ric1} R. W. Richardson, Phys. Lett. 3, 277 (1963); J. Math. Phys. 6, 1034 (1965).

\bibitem{ric2} R. W. Richardson and N. Sherman, Nucl. Phys. 52, 221 (1964).

\bibitem{vi1} P. Ribeiro, J. Vidal and R. Mosseri,
Phys. Rev. Lett. 99, 050402 (2007).

\bibitem{vi2} P. Ribeiro, J. Vidal and R. Mosseri,
Phys. Rev. E 78, 021106 (2008).

\bibitem{pan} F. Pan and J. P. Draayer, Phys. Lett. B 451, 1 (1999).

\bibitem{mori} H. Morita, H. Ohnishi, J. da Provid\^{e}cia and S. Nishiyama,
Nucl. Phys. B, 737, 337 (2006).

\bibitem{zhang} Y.-H Lee, J. Links and Y.-Z. Zhang, Nonlinearity 24, 1975 (2011).

\bibitem{mar} M. A. Marchiolli, D. Galetti and T. Debarba, Int. J. Quant. Inf. 11, 1330001 (2013).

\bibitem{pan1} F. Pan and J. P. Draayer, Ann. Phys. (N. Y.) 275, 224 (1999).

\bibitem{pan20169} F. Pan, Y.-Z. Zhang and J. P. Draayer,
Exact solution of the two-axis countertwisting Hamiltonian, arXiv: 1609.05581.

\bibitem{3} F. Pan, L. Bao, L. Zhai, X. Cui and J. P. Draayer, J. Phys. A {44}, 395305 (2011).

\bibitem{4} X. Guan, K. D. Launey, M. Xie, L. Bao, F. Pan and J. P. Draayer,
Phys. Rev. C {86}, 024313 (2012).

\bibitem{5} F. Pan, B. Li, Y.-Z. Zhang and J. P. Draayer,  Phys. Rev. C {88}, 034305 (2013).

\bibitem{6} X. Guan, K. D. Launey, M. Xie, L. Bao, F. Pan and J. P. Draayer,
Comp. Phys. Commun. {185},  2714 (2014).

\bibitem{rr1} A. Faribault, O. El Araby, C. Str\"{a}ter, V. Gritsev, Phys. Rev. B 83 (2011) 235124.

\bibitem{rr2} O. El Araby, V. Gritsev, A. Faribault, Phys. Rev. B 85 (2012) 115130.

\bibitem{1509} M. Bhattacharya, Analytical solvability of the two-axis countertwisting spin squeezing Hamiltonian,
arXiv:1509.08530.

\bibitem{rev67} T. D. Thomas and J. R. Grover, Phys. Rev. 159, 980 (1967).

\end{thebibliography}
\end{document}

\begin{appendix}
{\section{${\rm\large\bf Appendix:~ The~ general~ form ~of~ the~ Hamiltonian ~after ~the~transformations}$\label{A}}}
\vskip .3cm
As shown in (\ref{TA2}), the Hamiltonian after the Jordan-Schwinger realization (\ref{2})
and the variable substitution  is given by

\begin{eqnarray}\label{A1}
{H}_{\rm TA}={\chi\over{i}}\left( (1+2\delta_{\hat{\nu}_{b}1})  c^{\dagger}d-(1+2\delta_{\hat{\nu}_{a}1})d^{\dagger}c
 +2c^{\dagger}d^{\dagger}(d^{2}-c^{2}) \right).
 \end{eqnarray}

In general, we can make the following non-unitary transformation:

 \begin{eqnarray}\label{a0}\nonumber
 a^{\dagger}_{1}=\sqrt{\alpha\over{2}}c^{\dagger}+i \sqrt{\beta\over{2}}d^{\dagger},~~\tilde{a}_{1}=\sqrt{1\over{2\alpha}}c-i \sqrt{1\over{2\beta}}d,\\
 a^{\dagger}_{2}=\sqrt{\alpha\over{2}}c^{\dagger}-i \sqrt{\beta\over{2}}d^{\dagger},~~\tilde{a}_{2}=\sqrt{1\over{2\alpha}}c+ i \sqrt{1\over{2\beta}}d,
  \end{eqnarray}
where $\alpha=1+2\delta_{\hat{\nu}_{b}1}$ and $\beta=1+2\delta_{\hat{\nu}_{a}1}$,
which satisfy commutation relations

\begin{equation}
[\tilde{a}_{l},~a^{\dagger}_{j}]=\delta_{lj},~~[\tilde{a}_{l},~\tilde{a}_{j}]=0,~~[a^{\dagger}_{l},~a^{\dagger}_{j}]=0,
\end{equation}
but $\left(\tilde{a}_{j}\right)^{\dagger}\neq a_{j}$.
 the Hamiltonian (\ref{A1}) can then be expressed as
\begin{equation}\label{A2}
 \hat{H}_{\rm TA}={\chi}\left( \sqrt{\alpha\beta}( a^{\dagger}_{1}\tilde{a}_{1}-a^{\dagger}_{2}\tilde{a}_{2})
 +{1\over{2\sqrt{\alpha\beta}}}(a^{\dagger 2}_{1}-a^{\dagger 2}_{2})((\alpha+\beta)(\tilde{a}^{2}_{1}+\tilde{a}^{2}_{2})+
 2(\alpha-\beta)\tilde{a}_{1}\tilde{a}_{2}
 )\right).
 \end{equation}

When $\alpha\neq\beta$, which occurs in half-integer $J$ cases,
the new ingredient in (\ref{A2}) is the last term with new boson pairing operator $\tilde{a}_{1}\tilde{a}_{2}$.
Besides $S_{+}(w)$, it seems that an additional pairing operator $a^{\dagger}_{1}a^{\dagger}_{2}$ should also be needed in
the Bethe ansatz. However, the first term in (\ref{A2}) is commutative with $a^{\dagger}_{1}a^{\dagger}_{2}$,
namely, $[  a^{\dagger}_{1}\tilde{a}_{1}-a^{\dagger}_{2}\tilde{a}_{2},~a^{\dagger}_{1}a^{\dagger}_{2}]=0$,
while

\begin{eqnarray}\label{A3}\nonumber
&[\tilde{a}_{1}\tilde{a}_{2},~S_{+}(w)]={1\over{1-w}} a^{\dagger}_{1}\tilde{a}_{2}+
{1\over{1+w}} a^{\dagger}_{2}\tilde{a}_{1},~~
[\tilde{a}_{1}\tilde{a}_{2},~a^{\dagger}_{1}a^{\dagger}_{2}]=a^{\dagger}_{1}\tilde{a}_{1}+a^{\dagger}_{2}\tilde{a}_{2}+1,\\\nonumber
&[[\tilde{a}_{1}\tilde{a}_{2},~S_{+}(w)],~a^{\dagger}_{1}a^{\dagger}_{2}]=S_{+}(w),~~
[[\tilde{a}_{1}\tilde{a}_{2},~a^{\dagger}_{1}a^{\dagger}_{2}],~S_{+}(w)]=2S_{+}(w),\\
&[[\tilde{a}_{1}\tilde{a}_{2},~S_{+}(w)],~S_{+}(w^{\prime})]=
\left({1\over{(1-w)(1+w^{\prime})}} +{1\over{(1+w)(1-w^{\prime})}}
\right)a^{\dagger}_{1}a^{\dagger}_{2}.
\end{eqnarray}
It is the last commutation relation shown in (\ref{A3}) that makes
the Bethe ansatz solution rather complicated.

\end{appendix}